\documentclass[twocolumn,floatfix,aps,pre]{revtex4}
\usepackage{epsfig}
\usepackage{amsmath}
\usepackage{times}
\usepackage{graphicx}

\begin{document}

\title{Reptational dynamics in dissipative particle dynamics simulations of
       polymer melts}

\author{Petri Nikunen}
\affiliation{Biophysics and Statistical Mechanics Group, 
Laboratory of Computational Engineering, Helsinki University of Technology, 
P.O. Box 9203, FI-02015 HUT, Finland}

\author{Ilpo Vattulainen}
\affiliation{\mbox{Laboratory of Physics and Helsinki Institute of Physics, 
Helsinki University of Technology, P.O. Box 1100,} 
\mbox{FI-02015 HUT, Finland; MEMPHYS--Center for 
Biomembrane Physics, University of Southern Denmark,} 
\mbox{Odense, Denmark; Institute of Physics, Tampere 
University of Technology, P.O. Box 692, FI-33101 Tampere, Finland}}

\author{Mikko Karttunen}
\affiliation{\mbox{Biophysics and Statistical Mechanics Group, 
Laboratory of Computational Engineering,} 
\mbox{Helsinki University of Technology, 
P.O. Box 9203, FI-02015 HUT, Finland;} 
\mbox{Department of Applied Mathematics,
University of Western Ontario, 
London (ON), Canada}}

\date{November 27, 2005}

\begin{abstract}
Understanding the complex viscoelastic properties of polymeric liquids
remains a challenge in materials science and soft matter physics. 
Here, we present a simple and computationally efficient criterion for 
the topological constraints in polymeric liquids using the Dissipative 
Particle Dynamics (DPD). The same approach is also applicable in other 
soft potential models. For short chains the model correctly reproduces 
Rouse-like dynamics whereas for longer chains the dynamics becomes 
reptational as the chain length is increased---something that is not 
attainable using standard DPD or other coarse-grained soft potential 
methods. Importantly, no new length scales or forces need to be added.
\end{abstract}

\maketitle

\section{Introduction} 

The static and dynamic properties of polymeric liquids are, by and large,
dominated by topological constraints. The origin of these constraints is 
easy to understand: polymers can slide past but not penetrate through each other. That is the physical origin of the reptation
model~\cite{Edwards:67bf,Gennes:71ch,Doi:86ud} which is the most 
successful theory in describing the behavior of entangled polymers.
Despite active research in the field, entangled
polymeric liquids keep posing many challenges to 
theorists~\cite{McLeish:02su,Everaers:04dg,Sukumaran:05fe}, 
experimentalists~\cite{Richter:90jg,Kas:94gy,Schleger:98qt} and 
computational 
modelers~\cite{Kremer:88oj,Putz:00aa,Paul:91je,Padding:01wp,Paul:04uv,Ottinger:04cw,Kremer:05ud}. 
The importance of understanding the fundamentals of
polymeric liquids can hardly be overemphasized as they are 
one of the key issues in novel (bio)materials science~\cite{Forster:02as,Rastogi:05zs}.
%An extensive review of the properties of entangled polymeric systems
%is given by McLeish~\cite{McLeish:02su}.

The dynamics of polymer melts is typically described in terms of the Rouse 
and reptation models~\cite{Doi:86ud}. Short chains are able to
move to any direction and are not entangled. That is the physical origin of
 the Rouse model~\cite{Rouse:53jh,Doi:86ud}. For longer chains, entanglements 
 and uncrossability of chains cannot be ignored, and the chains become constrained to 
 move in the direction of the chain backbone. This motion resembles that of a 
 reptating snake---hence the name reptation 
model~\cite{Edwards:67bf,Gennes:71ch,Doi:86ud}.  

Computer simulations offer a detailed look into polymers and their dynamics. 
In classical molecular dynamics simulations the system size and simulation 
time pose limits as they are typically of the order of 10\,nm in linear 
size and around 10\,ns in time. In contrast, coarse grained methods, such
as dissipative particle dynamics  (DPD),
allow access to micrometer and microsecond scales. That is due to the
soft potentials, and, like everything in life, they do not come without
a price to pay: the softness of the conservative potentials allows 
the chains to slide
through each other thus strongly affecting the dynamics of the system. 
Indeed, the scaling laws obtained from DPD simulations of polymer melts~\cite{Spenley:00ce,Guerrault:04px} are not able
to describe entangled liquids. This is a direct consequence of the fact that in DPD simulations, polymers can penetrate through themselves. 
Whereas that is not a problem in studying the equilibrium properties in
the Rouse regime, reptation cannot be studied using the basic DPD model 
with soft interactions.

To preserve the advantages of coarse-grained models and to correct for
their deficiencies, 
Padding and Briels\,\cite{Padding:01wp} recently introduced an algorithm that explicitly detects and prevents bond crossings. They consider bonds as elastic bands that become entangled and use energy minimization to determine the entanglement positions.  This approach is general and very promising 
but it is also complicated to implement and computationally 
intensive\,\cite{Padding:01wp}. 

Another promising approach was put forward by
Pan and Manke~\cite{Pan:03ys}.  
They reduce the frequency of bond crossings by introducing segmental repulsive forces between the points of nearest contact
between neighboring chains. This approach is simple to implement but the introduction of a new force and a related cutoff increases the computational load, and adds a new length scale whose physical determination is somewhat
ambiguous. On the other hand, the model seems to be able to capture both the
Rouse and reptational behavior~\cite{Pan:03ys} like that of Padding and Briels\,\cite{Padding:01wp,Padding:02vb}.

In this article, we introduce a simple and generic criterion based 
on simple geometrical arguments
to solve the crossability problem. No new forces are added, the approach is 
conceptually simple and does not depend on the level of coarse graining.
Importantly, it allows easy, and if necessary even on-the-fly, 
tuning between the Rouse and reptation regimes.

The rest of this paper is organized as follows. In the next section we
will briefly describe the DPD method. Section\,\ref{sec:model} describes
our criterion for including topological constraints in DPD, or for that matter
any other soft potential, simulation. In Sec.\,\ref{sec:results} we show 
results from our simulations and compare them to other methods. Finally,
we finish with a discussion and outlook in Sec.\,\ref{sec:discussion}

\section{Dissipative particle dynamics}

In DPD, the time evolution of particles is given by
the Newton's equations of motion, and
the total force acting on particle 
$i$ is given as a sum of pairwise conservative, dissipative, and random forces, respectively,  as
$
  \vec{F}_i = \sum_{i\neq j}
  ( \vec{F}_{ij}^C + \vec{F}_{ij}^D + \vec{F}_{ij}^R ). 
$

The conservative force is independent of the dissipative and random forces. 
Typically it takes the form 
  \begin{equation}
  \label{eq:fcij}
  \vec{F}_{ij}^C = \left\{
  \begin{array}{ll}
    a_{ij}(1 - r_{ij}/r_c)\vec{e}_{ij}, & \mbox{if $r_{ij}<r_c$} \\
    0, & \mbox{otherwise},
  \end{array}
  \right.
  \end{equation}
with $\vec{r}_{ij}\equiv\vec{r}_i-\vec{r}_j$, $r_{ij}\equiv|\vec{r}_{ij}|$, 
and $\vec{e}_{ij}\equiv\vec{r}_{ij}/r_{ij}$. The variable $a_{ij}$
describes the repulsion between particles $i$ and $j$, and thus produces 
excluded volume interactions.

The dissipative force is expressed as
  \begin{equation}
  \label{eq:fdij}
  \vec{F}_{ij}^D = 
  -\gamma\omega^D(r_{ij})(\vec{v}_{ij}\cdot\vec{e}_{ij})\vec{e}_{ij}, 
  \end{equation}
where $\gamma$ is a friction parameter, $\omega^D(r_{ij})$ a weight 
function for the dissipative force, 
and $\vec{v}_{ij}\equiv\vec{v}_i-\vec{v}_j$. The dissipative force 
slows down the particles by decreasing kinetic energy from them. 
This effect is balanced by the random force due to thermal fluctuations,
  \begin{equation}
  \label{eq:frij}
  \vec{F}_{ij}^R = \sigma\omega^R(r_{ij})\zeta_{ij}\vec{e}_{ij},
  \end{equation}
where $\sigma$ is the amplitude of thermal noise,
$\omega^R(r_{ij})$ is the weight 
function for the random force, 
and $\zeta_{ij}(t)$ are Gaussian random variables with
$\langle\zeta_{ij}(t)\rangle = 0$ and 
$\langle\zeta_{ij}(t)\zeta_{kl}(t')\rangle =
(\delta_{ik}\delta_{jl}+\delta_{il}\delta_{jk})\delta(t-t')$.
The condition $\zeta_{ij}(t)=\zeta_{ji}(t)$ is required for momentum 
conservation. That is a necessary condition for the 
conservation of hydrodynamics.

The weight functions $\omega^D(r_{ij})$ and $\omega^R(r_{ij})$ cannot be 
chosen arbitrarily. Espa\~nol and Warren \cite{Espanol:95nr} 
showed that the fluctuation-dissipation relations 
$\omega^D(r_{ij}) = [\omega^R(r_{ij})]^2$Êand $\sigma^2 = 2\gamma k_BT$
must be satisfied for the system to have a canonical equilibrium distribution.
Here $T$ is the temperature of the system and $k_B$ is the 
Boltzmann constant. The functional form of the weight functions is not 
defined by the DPD method but virtually all DPD studies use
  \begin{equation}
  \omega^D(r_{ij}) = [\omega^R(r_{ij})]^2 = \left\{
  \begin{array}{ll}
    (1 - r_{ij}/r_c)^2, & \mbox{if $r_{ij}<r_c$} \\
    0, & \mbox{otherwise}.
  \end{array}
  \right.
  \end{equation}

Coarse graining in DPD comes in through the soft conservative potential and
forces (Eq.~(\ref{eq:fcij})). A detailed account of DPD, derivation of 
time and length scales, and its applications is given by R.D. Groot~\cite{Groot:04al}. An in depth discussion of coarse graining by
P. Espa{\~n}ol can be found in the same reference.

\section{Topological constraints \label{sec:model}}

\begin{figure}
\centering\epsfig{figure=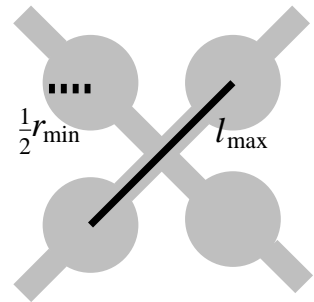,width=4cm}
\caption{Two bonds crossing each other. If Eq. (\ref{eq:bonds}) is 
satisfied, crossings cannot occur.}
\label{fig:bonds}
\end{figure}

To take into account the topological constraints, chain crossings
must be prevented. As discussed in the introduction,
there are currently two off-lattice methods~\cite{Padding:01wp,Pan:03ys}
for this purpose. Here, we introduce a third alternative. 

First, each individual bead has a radius $r_{\rm min}$ which is impenetrable
to other beads. In systems with Lennard-Jones potentials that condition is
automatically satisfied due to the $r^{-12}$ part that takes care of the 
Fermi exclusion principle. In mesoscopic simulations with soft potentials 
that constraint needs special attention. Second, the intramolecular bonds have some maximum
stretch, $\ell_{\rm max}$. 
By using simple geometry,
we can postulate that if the condition 
    \begin{equation}
    \label{eq:bonds}
    \sqrt{2} r_{\rm min} > \ell_{\rm max}
    \end{equation}
is satisfied, any two bonds cannot cross each other, see Fig.~\ref{fig:bonds}. 
The length scales involved, i.\,e., $r_{\rm min}$ and $\ell_{\rm max}$
have a clear physical meaning.

The obvious question is whether that condition is actually useful and when
does it work. As an example, let us consider DPD simulations of 
polymers. The parameters used in these simulations are often 
justified on the basis of the Flory-Huggins theory 
\cite{Groot:97xm,Groot:04al}, where the 
key component is the solubility as expressed by the 
$\chi$-parameters. Then, in simulations of block co-polymers, e.g., 
it is the mutual repulsion between the different components that 
matters---as a matter of fact, the values of the interaction 
parameters $a_{ij}$ may be derived in different ways and
their values tell only about the degree of coarse-graining. 
The condition set by Eq.~(\ref{eq:bonds}) can thus be met by 
a proper degree of coarse graining, complemented by a reasonable 
description for bond stretching (springs). Indeed, 
above $\ell_{\rm max}$ is limited by the type of springs used
in the model. With FENE springs~\cite{Grest:86ig} that is easy to
tune as they have only finite extension after which the force becomes
infinite.  With harmonic springs
more care is needed to satisfy Eq.~(\ref{eq:bonds}) as there is
no FENE-like cutoff set by the equation of motion. We will return to that
in the Sec.~\ref{sec:results}.

\section{Simulations  \label{sec:simu}}

For simplicity, and to be able to compare the model with other simulations, 
we considered a melt of linear polymers in a cubic box (3D) with periodic
boundary conditions. To avoid finite size effects, the linear
box size $L$ was chosen to be at least 1.75 times the average 
end-to-end distance of chains. 
We also carried out simulations with different box sizes to ensure
that the systems were free of finite size effects. That
was done since it is known that static properties are
affected by them~\cite{Kremer:90dj} .
%The results were the same within the statistical errors.

All the systems had 128 chains consisting of $N$ monomers, and no
additional solvent or free monomers were present. 
All monomers were chosen to be 
identical, and thus the monomer mass was set equal to unity, $m = 1$, 
fixing the scale of mass. 
The cutoff distance $r_c$ sets the length 
scale for the model.  
The conservative forces had the form given in Eq.~(\ref{eq:fcij}), with $r_c=1$ 
and $a_{ij}=a$ for all particle pairs. The values of $a$,
as well as other simulation parameters used in these
simulations are listed in Table~\ref{table:param}. 

For the random and dissipative forces we used Eqs.~(\ref{eq:fdij}) and (\ref{eq:frij}) with
the common choices~\cite{Groot:97xm,Groot:04al,Besold:00wp,Shillcock:05gw}
$\gamma=4.5$ and $\sigma=3$. This sets the temperature to $k_BT=1$, 
and hence the time scale is given by 
$\sqrt{m r_c^2 / k_B T}$.
%%% {\bf (CHECK THIS ONE!)}.

The monomers were connected using harmonic springs, i.\,e.,
%
%  \begin{equation}
$
  \vec{F}_i^S = \sum_j k ( \ell - r_{ij} ) \vec{e}_{ij},
$
%  \end{equation}
%
where the sum runs over all particles $j$ to which particle $i$ is connected. 
The equilibrium bond length was set to $\ell=0.95$. That particular value
was chosen as it is very near the first maximum of the radial 
distribution function (at the density $\rho=1$). 
The spring constant was chosen to be $k=2a$. If $k$ is much smaller, bonds are 
very flexible and Eq.~(\ref{eq:bonds}) is not satisfied. On the other hand, 
if $k$ is much larger than $a$, the time step $\Delta t$ must be decreased from 
the value set by the choice of $a$ thus decreasing the computational efficiency.
Another possibility would be to use FENE springs~\cite{Kremer:90dj} since
they have finite extension.

\begin{table}
\label{table:param} 
%\begin{tabular}{l@{\quad}|@{\quad}l@{\quad}l@{\quad}l}
\begin{tabular}{|l|l|l|l|l|}
\hline\hline
$a$ (amplitude of the conservative force)        &  50    & 100   &  150    &  200     \\
$k$ (spring constant)        &  100   & 200   &  300    &  400     \\
$\Delta t$  (time step) &  0.03  & 0.02  &  0.015  &  0.0125  \\
\hline\hline
\end{tabular}
\caption{The parameters used in this study.
}
\end{table}

The density was chosen to be $\rho=1$, which is lower than 
the densities typically used in DPD simulations~\cite{Groot:97xm,Groot:04al}. 
The reason for high 
densities is to give different repulsive interactions for different 
particle types. This works only if particles overlap each other 
considerably. In the present work, we don't need such interactions, and 
therefore the lower density is sufficient. In fact, the density of 
$\rho=1$ sets the monomer-monomer coordination number near 12, which 
is a typical value for real liquids.

All systems were started from random flight initial configurations and they
were equilibrated for $10^6$ time steps.
After the equilibration, we simulated systems at least for $10^7$ time steps to compute the desired quantities. 
Equations of motion were integrated using
the DPD-VV algorithm~\cite{Besold:00wp,Vattulainen:02bp,Nikunen:03db}.

\section{Results \label{sec:results}}

\begin{figure}
\includegraphics[width=0.22\textwidth,clip=true,viewport=8 0 127 120]
{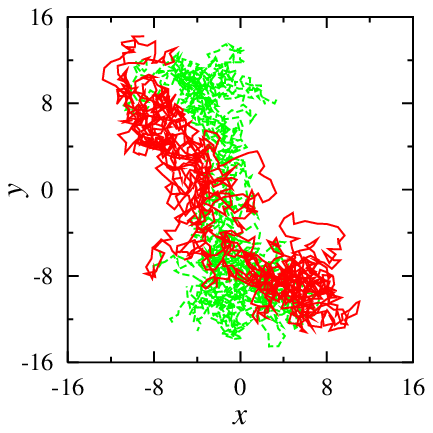}
\hfill
\includegraphics[width=0.22\textwidth,clip=true,viewport=8 0 127 120]
{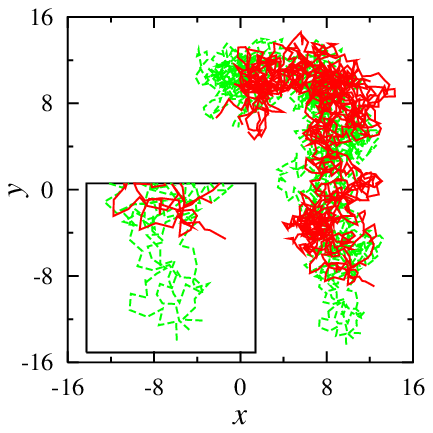}\\
\hspace*{-3cm}\textbf{a)} \hspace*{5cm}\textbf{b)}
\caption{The snapshots present 10 superpositions of configurations
for a chain of length $N=256$ 
taken at times 100 apart. Green: times up to 500 (in DPD time units), red:
times from 500 to 1000.
\textbf{a)} Rouse dynamics ($a=25$, $k=50$)
and
\textbf{b)}  reptation ($a=100$, $k=200$). 
}
\label{fig:snapshot}
\end{figure}

Figure~\ref{fig:snapshot} shows snapshots of the chain motion during 
the simulation at different times and regimes. For clarity, the
chain is projected onto two dimensions. It is immediately clear that
the motions in Figs.~\ref{fig:snapshot}a and ~\ref{fig:snapshot}b
are qualitatively different.
Figure~\ref{fig:snapshot}a shows Rouse-like motion in which the 
polymers are free to move in every direction, and 
Fig.~\ref{fig:snapshot}b represents reptation confined into a tube.

\subsection{Radial distribution function}

We will now study the static properties to see the effect of 
Eq.~(\ref{eq:bonds}). As discussed, by tuning the chain stiffness
it is possible to move gradually from the Rouse regime to reptation.
This should be reflected in both the radial distribution function
and the bond length distribution. 

\begin{figure}
%\centering
\includegraphics[width=0.235\textwidth,clip=true,viewport=2 0 121 120]
{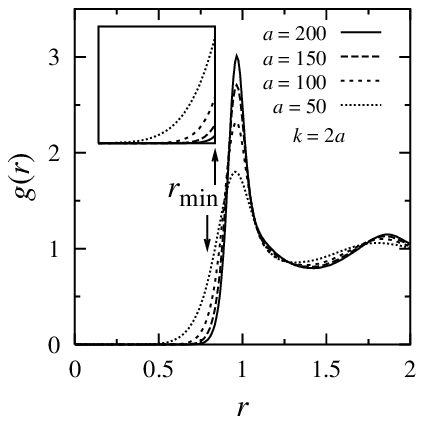}
\hfill
\includegraphics[width=0.235\textwidth,clip=true,viewport=2 0 122 120]
{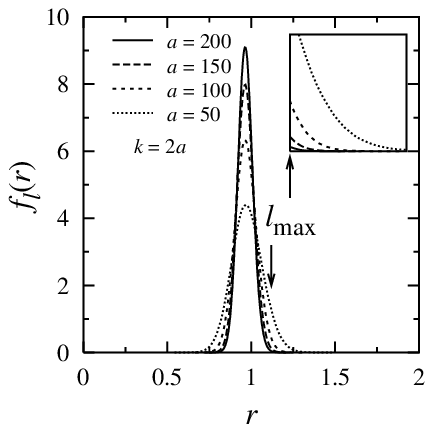}\\
\hspace*{-3cm}\textbf{a)} \hspace*{5cm}\textbf{b)}
\caption{\textbf{a)} Radial distribution function in the case of $N=32$. 
The arrow shows the distance $r_{\rm min}$, and the inset the
region at lengths shorter than $r_{\rm min}$. 
Compare with LJ models ($r_c=1,2.5$).
\textbf{b)} Bond length distribution ($N=32$). The arrow shows the 
location of $\ell_{\rm max}$, and the inset the region at values
larger than $\ell_{\rm max}$. 
}
\label{fig:rdffl}
\end{figure}

The radial distribution function (RDF) $g(r)$ describes the qualitative 
structure of a fluid. It is defined as
$
  g(r) = \rho(r)/\rho 
$
where $\rho(r)$ is the average density from a given 
particle at a distance $r$.
Figures~\ref{fig:rdffl}a and \ref{fig:rdffl}b show the radial 
distribution function 
$g(r)$ and the bond length distribution $f_\ell(r)$ for different 
parameter sets for chains of length $N=32$. 
The arrows in the figures indicate the values of $r_{\rm min}$ 
and $\ell_{\rm max}$ in Eq.~(\ref{eq:bonds}). 
As the figures show, Eq.~(\ref{eq:bonds}) is satisfied for 
larger values of $a$ and $k$. The small non-zero values below
$r_{\rm min}$ are due to the softness of the interparticle
DPD potentials.  

The above can be characterized by taking a look at the average bond
lengths are their mean square deviations. For $a_{ij}=50$ 
we measured $\langle \ell \rangle = 0.977 \pm 0.092$. As 
the strength of interaction is increased we obtain
$\langle \ell \rangle = 0.968 \pm 0.064$ for $a_{ij}=100$,
$\langle \ell \rangle = 0.966 \pm 0.051$ for $a_{ij}=150$,
and 
$\langle \ell \rangle = 0.965 \pm 0.045$ for $a_{ij}=200$.
The most important issue is the decrease of the mean square deviation
as that restricts the amount of overlap between the monomers of different
chains. Importantly, for FENE chains this can be directly controlled by 
using the above measurements and RDF as a guideline and setting the 
maximum extent of the chain to an appropriate value. 

A comparison of the radial distribution functions shows that the current
approach allows for tuning between typical 
DPD results~\cite{Groot:97xm,Nikunen:03db,Guerrault:04px} 
and typical molecular
dynamics simulations using Lennard-Jones potentials~\cite{Kremer:90dj}. 
As the bond strength is increased ($a\ge 100$, $k \ge 200$), $g(r)$ becomes 
qualitatively similar to that from a Lennard-Jones system.

\subsection{Static scaling}

\begin{figure}
\includegraphics[width=0.23\textwidth,clip=true,viewport=2 0 130 120]
{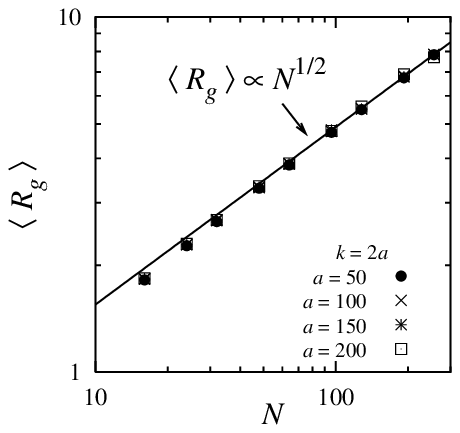}
\hfill
\includegraphics[width=0.23\textwidth,clip=true,viewport=11 0 138 120]
{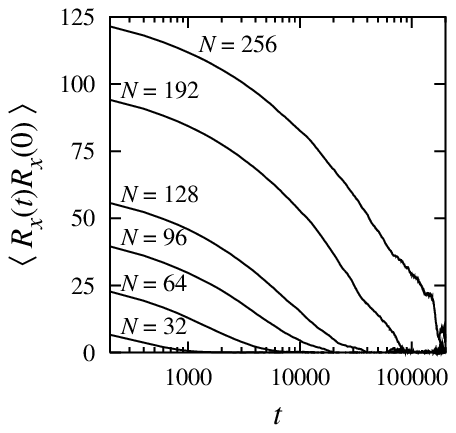}\\
\hspace*{-3cm}\textbf{a)} \hspace*{5cm}\textbf{b)}
\caption{\textbf{a)} The radius of gyration as a function of chain length
for different parameters.
\textbf{b)} The end-to-end vector autocorrelation ($a=100$, $k=200$)
for chains of different length.
}
\label{fig:rga0}
\end{figure}

Next, we studied the end-to-end distance $R$ 
and the radius of gyration 
$R_g$. The former is defined as 
$R = |\vec{R}| = | \vec{r_1}-\vec{r_N}|$ and the latter as
$R_g^2 = \frac{1}{N} \sum_{i=1}^N ( \vec{r}_i-\vec{r}_{\rm cm} )^2$,
where $\vec{r}_{\rm cm}=\frac{1}{N}\sum_{i=1}^N\vec{r}_i$. In 
a polymer melt, they are expected to scale as
%
%  \begin{eqnarray}
%  \langle R \rangle &\propto& N^{1/2}; \\
%  \langle R_g \rangle &\propto& N^{1/2}.
%  \end{eqnarray}
$
\langle R \rangle \propto N^{1/2} \mathrm{\,\,\, and \,\,\,}
\langle R_g \rangle \propto N^{1/2}
$
in both Rouse and the reptation regime. Previous studies using soft 
potentials~\cite{Spenley:00ce} and systems with more 
realistic hard potentials~\cite{Kremer:90dj} exhibit
scaling. 
In Fig.~\ref{fig:rga0}a we plot the results for the radius of
gyration for different parameter sets. It is clear from the figure
that the system exhibits the proper scaling behavior 
independently of the interaction parameters as it should.

\subsection{Relaxation time}

One of the main practical obstacles in simulations of polymeric solutions
is the long stress relaxation time. The longest relaxation time, $\tau$, 
depends on
the molecular weight and the reptation theory predicts it to scale
as $\tau\propto N^{3}$. That prediction assumes only one mechanism for
relaxation, i.\,e., diffusion along the 
countour~\cite{Edwards:67bf}.
The Rouse model predicts a distinctly different behavior with 
$\tau\propto N^{2}$.

\begin{figure}
\includegraphics[width=0.235\textwidth,clip=true,viewport=2 0 135 120]
{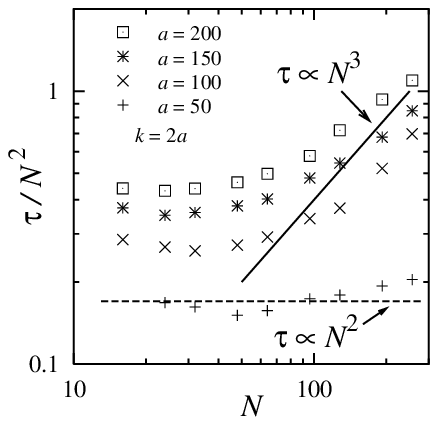}
\hfill
\includegraphics[width=0.235\textwidth,clip=true,viewport=9 0 142 120]
{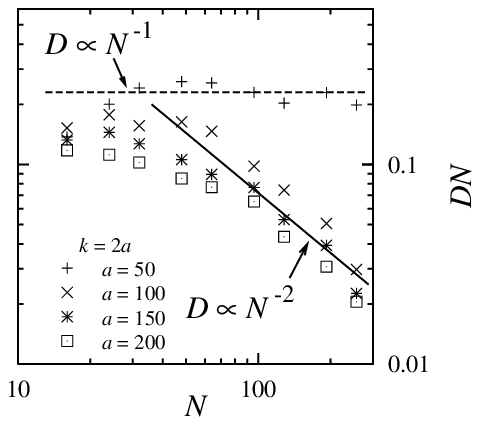}\\
\hspace*{-3cm}\textbf{a)} \hspace*{5cm}\textbf{b)}
\caption{\textbf{a)} Scaling of the longest relaxation time $\tau$. 
There is a crossover
from Rouse scaling ($\tau \propto N^2$) to reptation ($\tau \propto N^3$).
\textbf{b)} Similarly, the proper scaling limits are reached for the 
diffusion coefficient $D$.
}
\label{fig:tauD}
\end{figure}

To estimate the scaling behavior, we measured the end-to-end
autocorrelation function. It is shown in  
Fig.~\ref{fig:rga0}b for polymers of  
different length. Assuming exponential decay, i.\,e., 
$$
  \langle \vec{R}(t) \cdot \vec{R}(0) \rangle \sim \exp (-t/\tau),
$$
we can extract the longest relaxation time $\tau$ by fitting. 
Figure~\ref{fig:tauD}a shows that both scaling regimes are captured properly. 
Fig.~\ref{fig:tauD}a
illustrates one of the main results of this paper: the simple 
criterion summarized by Eq.~(\ref{eq:bonds})  allows an easy,
physical  
and computationally efficient tuning between the Rouse regime
and reptation. 

The scaling exponents 2 and 3 for Rouse and reptation, 
in respective order, are the 
limiting laws. The exponents have been frequently debated 
in the literature.
For example, 
for the same values of $N$ as used here, Padding and Briels\,\cite{Padding:02vb} 
found two scaling regimes in $\tau$, one with exponent $2.8$ and the other one with exponent $3.5$. 
The dependence $\tau\propto N^{3.4}$ is experimentally observed for the longest relaxation time in 
the entangled regime~\cite{Doi:81ac}. 
This discrepancy is often associated with 
fluctuations of the contour length of the primitive path. In a real situation, 
however, the tube has a characteristic lifetime and the length of the primitive path 
fluctuates since the Rouse modes continue in the direction along the primitive path. 

Determining the value of the exponent was not the main goal here, and thus we
did not attempt to evaluate it in a high precision---we  will focus on that in a future publication. 
The above simply to demonstrates that 
it is indeed possible to use the soft DPD model to describe 
entangled polymer liquids realistically without introducing additional
length scales and forces.

\subsection{Diffusion}

The motion of a polymer, or its segments, is described by the 
diffusion coefficient. Typically, one measures 
the center-of-mass diffusion coefficient for a polymer chain, i.\,e.,

$$
  D = \lim_{t\to\infty} \frac{1}{6} \langle \left[ 
  \vec{r}_{\rm cm}(t) - \vec{r}_{\rm cm}(0) \right]^2 \rangle.
$$

The scaling of $D$ with molecular weight has been studied
intensely over the years, see e.\,g. the discussion in~\cite{McLeish:02su}. 

The theory predicts two scaling limits,
$D \propto N^{-1}$ for the Rouse model 
and $D \propto N^{-2}$ for the pure reptation model 
($N$ is proportional to molecular weight).
Figure~\ref{fig:tauD}b shows that both scaling regimes are found. 

Considering the nature of the DPD model, it is remarkable that both 
regimes are recovered. In simulations using the plain DPD model without
paying attention to the criterion given by Eq.~(\ref{eq:bonds}), 
only Rouse scaling has been observed~\cite{Spenley:00ce,Guerrault:04px}
even in the case of long polymers. 

As with the longest relaxation time, the exponents typically reported
are between the scaling limits. 
Pearson {\it et al.}~\cite{Pearson:87su} measured $D$ as a function of molecular 
weight $M_w$ in polyethylene and they found that the diffusion coefficient follows a power law $D=1.65M_w^{-1.98}cm^2/s$ for the entire range from $M_w=600$ to $M_w=12000$ (g/mol). The simulations by Kremer 
\textit{et al.}~\cite{Kremer:90dj,Putz:00aa} and Padding and 
Briels\,\cite{Padding:01wp,Padding:02vb} confirmed this finding: the center of mass 
diffusion coefficient 
scales as $D\propto N^{-2}$ in melt. 

Padding and Briels\,\cite{Padding:02vb} compared their results to different 
simulations and experiments, and found that in ethylene the crossover between 
Rouse-like and reptational dynamics takes place at molecular weight of 560\,g/mol 
(corresponds to 40 ethylenes). Because in Fig.~\ref{fig:tauD}b the crossover takes 
place between $N=40$ and $N=60$, we can picture each particle roughly as one ethylene unit.

\section{Discussion \label{sec:discussion}}

In this article we have presented a simple  
criterion for topological constraints in coarse grained
DPD simulations of polymeric liquids. No new forces or length scales
were added.
We showed that this approach is able to reproduce the
Rouse model at one limit and reptational dynamics at the other.
Here, we validated and demonstrated this approach against other models
and experimental results using linear homopolymers.  
This approach can also be used for systems of, e.\,g.,
block co-polymers with different interactions and monomer sizes,
and shear simulations. In practice, one can always run a short test simulation,
and use $g(r)$ and the bond length distribution 
(as in Figs.~\ref{fig:rdffl}a and b) to verify that the 
criterion set by Eq.~(\ref{eq:bonds}) is met.

There is one other issue that we need to address, namely 
by tuning the chain stiffness one inevitably changes the entanglement length in addition
to intercrossability of chains. It  is known from previous simulations using Lennard-Jones
as well as some coarse-grained models that increasing chain stiffness intensifies 
reptation~\cite{Faller:00kr,Faller:01gn,Faller:02gz}.  
To account for this in order to use coarse-grained methods such as DPD with
soft potentials in a controlled way, one should use (at least) the persistence length as measure that 
should be matched between the coarse-grained and the atomistic models. 
Here, we did not attempt to do that systematically.

Here, we made no attempt to determine the precise scaling exponents
for the diffusion coefficient or the longest relaxation time. Though, there
are a lot of subtleties, such as the tube dimensions, lifetime, friction 
and the plateau
modules, related to the scaling behavior~\cite{Putz:00aa}. A future
publication will focus on them and the detailed mechanisms.

\begin{acknowledgments}

This work has been supported by Emil Aaltonen foundation (M.\,K.), 
the Academy of Finland through its
Center of Excellence Program (I.V.) and the Academy of Finland Grants 
Nos. 54113, 00119 (M.K.), and 80246 (I.V.). 
We also thank the Finnish IT Center for Science (CSC) and the HorseShoe (DCSC) 
supercluster computing facility at the University of Southern Denmark for 
computer resources.

\end{acknowledgments}

\end{document}